# Spatial Clustering of Galaxies in Large Datasets

Alexander S. Szalay, Department of Physics and Astronomy, Johns Hopkins University
Tamás Budavari, Department of Physics and Astronomy, Johns Hopkins University
Andrew Connolly, Department of Physics and Astronomy, University of Pittsburgh
Jim Gray, Microsoft Research
Takahiko Matsubara, Department of Physics and Astrophysics, Nagoya University
Adrian Pope, Department of Physics and Astronomy, Johns Hopkins University
István Szapudi, Institute of Astronomy, University of Hawaii

# Spatial Clustering of Galaxies in Large Datasets

Alexander S. Szalay[a], Tamás Budavari[a], Andrew Connolly[b], Jim Gray[c], Takahiko Matsubara[d], Adrian Pope[a] and István Szapudi[e]

[a]Department of Physics and Astronomy, The Johns Hopkins University, Baltimore, MD 21218
[b]Department of Physics and Astronomy, University of Pittsburgh, Pittsburgh, PA 15260
[c]Microsoft Research, San Francisco, CA 94105
[d]Department of Physics and Astrophysics, Nagoya University, Nagoya 464-8602, Japan
[e]Institute of Astronomy, University of Hawaii, Honolulu, HI 96822

## ABSTRACT

Datasets with tens of millions of galaxies present new challenges for the analysis of spatial clustering. We have built a framework that integrates a database of object catalogs, tools for creating masks of bad regions, and a fast (NlogN) correlation code. This system has enabled unprecedented efficiency in carrying out the analysis of galaxy clustering in the SDSS catalog. A similar approach is used to compute the three-dimensional spatial clustering of galaxies on very large scales. We describe our strategy to estimate the effect of photometric errors using a database. We discuss our efforts as an early example of data-intensive science. While it would have been possible to get these results without the framework we describe, it will be infeasible to perform these computations on the future huge datasets without using this framework.

## 1. INTRODUCTION

Developments in Astronomy detector size are improving exponentially. Consequently, Astronomy data volumes are more than doubling every year. This even exceeds the rate of Moore's law, describing the speedup of CPUs and growth of storage. This trend results from the emergence of large-scale surveys, like 2MASS, SDSS or 2dFGRS. Soon there will be almost all-sky data in more than ten wavebands. These large-scale surveys have another important characteristic: they are each done by a single group, with sound statistical plans and well-controlled systematics. As a result, the data are becoming increasingly more homogeneous, and approach a fair sample of the Universe. This trend has brought a lot of advances in the analysis of the large-scale galaxy distribution. Our goal today is to reach an unheard-of level of accuracy in measuring both the global cosmological parameters and the shape of the power spectrum of primordial fluctuations. The emerging huge data sets from wide field sky surveys pose interesting issues, both statistical and computational. One needs to reconsider the notion of optimal statistics. We discuss the power spectrum analysis of wide area galaxy surveys using the Karhunen-Loeve transform as a case study.

These large, homogenous datasets are also changing the way we approach their analysis. Traditionally, statistics in cosmology has primarily dealt with how to extract the most information from the small samples of galaxies we had. This is no longer the case: there are redshift surveys of 300,000 objects today; soon there will be a million measured galaxy redshifts. Angular catalogs today have samples in excess of 50 million galaxies; soon they will have 10 billion (LSST). In the observations of the CMB, COBE had a few thousand pixels on the sky, MAP will have a million, PLANCK will have more than 10 million. Thus, shot noise and sample size is no longer an issue. The limiting factors in these data sets are the systematic uncertainties, like photometric zero points, effects of seeing, uniformity of filters, etc.

The statistical issues are changing accordingly: it is increasingly important to find techniques that can be de-sensitized to systematic uncertainties. Many traditional statistical techniques in astronomy focused on `optimal' techniques. It was generally understood, that these minimized the statistical noise in the result, but they are quite sensitive to various systematics. Also, they assumed infinite computational resources. This was not an issue when sample sizes were in the thousands. But, many of these techniques involve matrix diagonalizations or inversions and so the computational cost scales as the 3rd power of matrix size. Samples a thousand times larger have computational costs a billion times higher.

Even if the speedup of our computers keeps up with the growth of our data, it cannot keep pace with with such powers. We need to find algorithms that scale more gently. In the near future we hypothesize that only algorithms with *NlogN* scaling will remain feasible.

As the statistical noise decreases with larger samples, another effect emerges: *cosmic variance*. This error term reflects the fact that our observing position is fixed at the Earth, and at any time we can only study a fixed – albeit ever increasing – region of the Universe. This provides an ultimate bound on the accuracy of any astronomical measurement. We should carefully keep this effect in mind when designing new experiments. In this paper we will discuss our goals, and the current state of the art techniques in extracting cosmological information from large data sets. In particular, we use the Karhunen-Loeve (KL) transform as a case study; showing how we had to do step by step improvements in order to turn an optimal method into a useful one.

**1.1 Motivation: precision cosmology**

We are entering the era of precision cosmology. The large new surveys with their well-defined systematics are key to this transition. There are many different measurements we can make that each constrain combinations of the cosmological parameters. For example, the fluctuations in the cosmic Microwave Background (CMB) around the multipole *l* of a few hundred are very sensitive to the overall curvature of the Universe, determined by both dark matter and dark energy (deBernardis et al 2001, Netterfield et al 2001).

Due to the expansion of the Universe, we can use redshifts to measure distances of galaxies. Since galaxies are not at rest in the frame of the expanding Universe, their motions cause an additional distortion in the line-of-sight coordinate. This property can be used to study the dynamics of galaxies, inferring the underlying mass density. Local redshift surveys can measure the amount of gravitating dark matter, but they are insensitive to the dark energy. Combining these different measurements (CMB + redshift surveys), each with their own degeneracy can yield considerably tighter constraints than either of them independently. We know most cosmological parameters to an accuracy of about 10% or somewhat better today. Soon we will be able to reach the regime of 2-5% relative errors, through both better data but also better statistical techniques.

The relevant parameters include the age of the Universe, $t_0$, the expansion rate of the Universe, also called Hubble's constant $H_0$, the deceleration parameter $q_0$, the density parameter $\Omega$, and its components, the dark energy, or cosmological constant $\Omega_L$, the dark matter $\Omega_m$, the baryon fraction $f_B$, and the curvature $\Omega_k$. These are not independent from one another, of course. Together, they determine the dynamic evolution of the Universe, assumed to be homogeneous and isotropic, described by a single scale factor $a(t)$. For a Euclidian (flat) Universe $\Omega_L+\Omega_m =1$. One can use both the dynamics, luminosities and angular sizes to constrain the cosmological parameters. Distant supernovae have been used as standard candles to get the first hints about a large cosmological constant. The angular size of the Doppler-peaks in the CMB fluctuations gave the first conclusive evidence for a flat universe, using the angular diameter-distance relation. The gravitational infall manifested in redshift-space distortions of galaxy surveys has been used to constrain the amount of dark matter.

These add up to a remarkably consistent picture today: a flat Universe, with $\Omega_L=0.7\pm0.05$, $\Omega_m=0.3\pm0.05$. It would be nice to have several independent measurements for the above quantities. Recently, new possibilities have arisen about the nature of the cosmological constant – it appears that there are many possibilities, like quintessence, that can be the dark energy. Now we are facing the challenge of coming up with measurements and statistical techniques to distinguish among these alternative models.

There are several parameters used to specify the shape of the fluctuation spectrum. These include the amplitude $\sigma_8$, the root-mean-square value of the density fluctuations in a sphere of 8 Mpc radius, the shape parameter $\Gamma$, the redshift-distortion parameter $\beta$, the bias parameter $b$, and the baryon fraction $f_B=\Omega_B/\Omega_m$. Other quantities, like the neutrino mass also affect the shape of the fluctuation spectrum, although in more subtle ways than the ones above (Seljak and Zaldarriega 1996). The shape of the fluctuation spectrum is another sensitive measure of the Big Bang at early times. Galaxy surveys have traditionally measured the fluctuations over much smaller scales (below 100 Mpc), where the fluctuations are nonlinear, and even the shape of the spectrum has been altered by gravitational infall and the dynamics of the Universe. The expected spectrum on very large spatial scales (over 200 Mpc) was shown by COBE to be scale-invariant, reflecting the primordial initial conditions, remarkably close to the predicted Zeldovich-Harrison shape. There are several interesting physical effects that will leave an imprint on the fluctuations: the scale of the horizon at recombination, the horizon at matter-radiation equality, and the sound-horizon—all between 100-200 Mpc (Eisenstein and Hu 1998). These scales have been rather difficult to measure: they used to be too small for CMB, too large for redshift surveys. This is rapidly changing, new, higher resolution CMB experiments are now covering sub-degree scales, corresponding to less than 100 Mpc comoving, and redshift surveys like 2dF and SDSS are reaching scales well above 300 Mpc.

We have yet to measure the overall contribution of baryons to the mass content of the Universe. We expect to find the counterparts of the CMB Doppler bumps in galaxy surveys as well, since these are the remnants of horizon scale fluctuations in the baryons at the time of recombination. The Universe behaved like a resonant cavity at the time. Due to the dominance of the dark matter over baryons the amplitude of these fluctuations is suppressed, but with high precision measurements they should be detectable. A small neutrino mass of a few electron volts is well within the realm of possibilities. Due to the very large cosmic abundance of relic neutrinos, even such a small mass would have an observable effect on the shape of the power spectrum of fluctuations. It is likely that the sensitivity of current redshift surveys will enable us to make a meaningful test of such a hypothesis. One can also use large angular catalogs, projections of a 3-dimensional random field to the sphere of the sky, to measure the projected power spectrum. This technique has the advantage that dynamical distortions due to the peculiar motions of the galaxies do not affect the projected distribution. The first such analyses show promise.

### 1.4 Large surveys

As mentioned in the introduction, some of the issues related to the statistical analysis of large redshift surveys, like 2dF (Percival et al 2001), or SDSS (York et al 2000) with nearly a billion objects are quite different from their predecessors with only a few thousand galaxies. The foremost difference is that shot-noise, the usual hurdle of the past is irrelevant. Astronomy is different from laboratory science because we cannot change the position of the observer at will. Our experiments in studying the Universe will never approach an ensemble average; there will always be an unavoidable *cosmic variance* in our analysis. By studying a larger region of the Universe (going deeper and/or wider) can decrease this term, but it will always be present in our statistics.

Systematic errors are the dominant source of uncertainties in large redshift surveys today. For example photometric calibrations, or various instrumental and natural foregrounds and backgrounds contribute bias to the observations. Sample selection is also becoming increasingly important. Multicolor surveys enable us to invert the observations into physical quantities, like redshift, luminosity and spectral type. Using these broadly defined *`photometric redshifts'*, we can select statistical subsamples based upon approximately rest-frame quantities, for the first time allowing meaningful comparisons between samples at low and high redshifts.

Various effects, like small-scale nonlinearities, or redshift space distortions, will turn an otherwise homogeneous and isotropic random process into a non-isotropic one. As a result, it is increasingly important to find statistical techniques, which can reject or incorporate some of these effects into the analysis. Some of these cannot be modeled analytically; we need to perform Monte-Carlo simulations to understand the impact of these systematic errors on the final results. The simulations themselves are also best performed using databases.

Data are organized into databases, instead of the flat files of the past. These databases contain several well-designed and efficient indices that make searches and counting much faster than brute-force methods. No matter which statistical analyses we seek to perform, much of the analysis consists of data filtering and counting. Up to now most of this has been performed off-line. Given the large samples in today's sky surveys, offline analysis is becoming increasingly inefficient – scientists want to be able to interact with the data. Here we would like to describe our first efforts to integrate large-scale statistical analyses with the database. Our analysis would have been very much harder, if not entirely infeasible, to perform on flat files.

# 2. STATISTICAL TECHNIQUES

The most frequent techniques used in analyzing data about spatial clustering are the two-point correlation functions and various power spectrum estimators. There is an extensive literature about the relative merits of each of the techniques. For an infinitely large data set in principle both techniques are equivalent. In practice however there are subtle differences, finite sample size affects the two estimators somewhat differently, edge effects show up in a slightly different fashion and there are also practical issues about computability and hypothesis testing, which are different for the two techniques.

## 2.1 Two-point correlation function

The most often used estimator for the two point correlations is the LS estimator (Landy and Szalay 1992)

$$x(r) = \frac{DD - 2DR + RR}{RR}$$

which has a minimal variance for a Poisson process. DD, DR and RR describe the respective normalized pair count in a given distance range. For this estimator and for correlation functions in general, hypothesis testing is somewhat cumbersome. If the correlation function is evaluated over a set of differential distance bins, these values are not independent, and their correlation matrix also depends on the three and four-point correlation functions, less familiar than the two-point function itself. The brute-force technique involves the computation of all pairs and binning them up, so it scales as $O(N^2)$. In terms of modeling systematic effects, it is very easy to compute the two-point correlation function between two points.

Another popular second order statistic is the power spectrum P(k), usually measured by using the FKP estimator (Feldman et al 1994). This is the Fourier-space equivalent of the LS estimator for correlation functions. It has both advantages and disadvantages over correlation functions. Hypothesis testing is much easier, since in Fourier space the power spectrum at two different wavenumbers are correlated, but the correlation is compact. It is determined by the *window-function*, the Fourier transform of the sample volume, usually very well understood. For most realistic surveys the window function is rather anisotropic, making angular averaging of the three-dimensional power spectrum estimator somewhat complicated. During hypothesis testing one is using the estimated values of P(k), either directly in 3D Fourier space, or compressed into quadratic sums binned by bands. Again, the 3rd and 4th order terms appear in the correlation matrix. The effects of systematic errors are much harder to estimate.

## 2.2 Hypothesis testing

Hypothesis testing is usually performed in a parametric fashion, with the assumption that the underlying random process is Gaussian. We evaluate the log likelihood as

$$\ln L(\Pi) = -\frac{1}{2} x^T C^{-1} x - \frac{1}{2} \ln |C|$$

where $x$ is the data vector, and $C$ is its correlation matrix, dependent on the parameter vector $\Pi$. There is a fundamental lower bound on the statistical error, given by the Fisher matrix, easily computed. This is a common tool used these days to evaluate the sensitivity of a given experiment to measure various cosmological parameters. For more detailed comparisons of these techniques see Tegmark et al (1998).

What would an ideal method be? It would be useful to retain much of the advantages of the 2-point correlations so that the systematics are easy to model, and those of the power spectra so that the modes are only weakly correlated. We would like to have a hypothesis testing correlation matrix without 3rd and 4th order quantities. Interestingly, there is such a method, given by the Karhunen-Loeve transform. In the following subsection we describe the method, and show why it is a useful framework for the analysis of the galaxy distribution. Then we discuss some of the detailed issues we had to deal with over the years to turn this into a practical tool.

One can also argue about parametric and non-parametric techniques, like using bandpowers to characterize the shape of the fluctuation spectrum. We postulate, that for the specific case of redshift surveys it is not possible to have a purely non-parametric analysis. While the shape of the power spectrum itself can be described in a non-parametric way, the distortions along the redshift direction are dependent on a physical model (gravitational infall). Thus, without an explicit parameterization or ignoring this effect no analysis is possible.

**2.3 Karhunen-Loeve analysis of redshift surveys**

The Karhunen-Loeve (KL) eigenfunctions (Karhunen 1947, Loeve 1948) provide a basis set in which the distribution of galaxies can be expanded. These eigenfunctions are computed for a given survey geometry and fiducial model of the power spectrum. For a Gaussian galaxy distribution, the KL eigenfunctions provide optimal estimates of model parameters, i.e. the resulting error bars are given by the inverse of the Fisher matrix for the parameters Vogeley & Szalay 1996). This is achieved by finding the orthonormal set of eigenfunctions that optimally balance the ideal of Fourier modes with the finite and peculiar geometry and selection function of a real survey. In this section we present the formalism for the KL analysis following the notation of Vogeley & Szalay (1996) who introduced this approach to galaxy clustering. The KL method has been applied to the Las Campanas redshift survey by Matsubara, Szalay & Landy (2000) and to the PSCz survey by Hamilton, Tegmark & Padmanabhan (2001).

The KL transform is often called *optimal subspace filtering* (Therrien 1992), describing the fact, that during the analysis some of the modes are discarded. This offers distinct advantages. If the measurement is composed of a signal that we want to measure (gravitational clustering) superposed on various backgrounds (shot-noise, selection effects, photometric errors, etc) which have slightly different statistical properties, the diagonalization of the correlation matrix can potentially segregate these different types of processes into their own subspaces. If we select our subspace carefully, we can actually improve on the signal to noise of our analysis.

The biggest advantage is that hypothesis testing is very easy and elegant. First of all, all KL modes are orthogonal to one another, even if the survey geometry is extremely anisotropic. Of course, none of the KL modes can be narrower than the survey window, and they shape is clearly affected by the survey geometry. The orthogonality of the modes represents repulsion between the modes, they cannot get too close; otherwise they could not be orthogonal. As a result the KL modes are dense-packed into Fourier-space, thus optimally representing the information enabled by the survey geometry.

Secondly, the KL transform is a linear transformation. If we do our likelihood testing over the KL-transform of the data, the likelihood correlation matrix contains only second order quantities. This avoids problems with 3 and 4 point functions. All these advantages became very apparent when we applied the KL method to real data.

## 3. WORKING WITH A DATABASE

**3.1 Why use a database**

The sheer size of the data involved makes it necessary to store the data in a database – there are just too many objects organize as directories of files. We originally started to use databases solely for this reason. Astronomers in general use databases only to store their data, but when they do science, they generate a flat file, usually in a simple table. Then they use their own code for the scientific analysis. Mostly the databases are remote. One has to enter queries into a web form, and retrieve the data as either and ASCII or binary table.

Some of us have been working on designing the Science Archive for the Sloan Digital Sky Survey. We are now using a relational database, Microsoft's SQL Server, as the back-end database. In order to efficiently perform queries that involve spatial boundaries, we have developed a class library based upon a hierarchical triangulation of the sky (Kunszt et al 2001) to handle searches over spherical polygons. We added the class library as an extension to SQL Server, so its functions can be called directly inside the database.

The database contains much more than just the basic photometric or spectroscopic properties of the individual objects. We have computed, as an add-on, the inversion of the photometric observations into physical, rest-frame parameters, like luminosity, spectral type, and of course a photometric redshift. Much of the training for this technique was obtained from the spectroscopic observations. These are stored in the database as an ancillary table. Information about the geometry of the survey, how it is organized into stripes, strips, runs, camera columns and fields, is also stored in the database. The value of seeing is monitored and saved in the Field table. The 'blind' pixels of the survey, caused by a bright star, satellite, meteorite or a ghost in the camera are also saved in the database as an extended object, with their convex hull and a bounding circle plus and a bounding rectangle.

We are using this database for subsequent scientific analysis. We have a local copy of the data at Johns Hopkins, stored in a relatively high performance, yet inexpensive database server. While building our applications to study the correlation properties of galaxies, we have discovered that many of the patterns in our statistical analysis involve tasks that are much better performed inside the database than outside, on flat files. The database gives high-speed sequential search of complex predicates using multiple CPUs, multiple disks, and large main memories. It also has sophisticated indexing and data combination algorithms that compete favorably with hand-written programs against flat files. Indeed, we see cases where multi-day batch file runs are replaced with database queries that run in minutes.

### 3.2 Generating statistical samples, masks

In order to generate meaningful results for the clustering, we need to create a well-defined, statistically fair sample of galaxies. We have to censor objects that are in areas of decreased sensitivity or bad seeing. We also have to be aware of the geometry of the censored areas. We created these 'masks' using plain database queries for fields of bad seeing, rectangles around bright stars, and other trouble spots. In the next release of the database these regions will be derived by processing images that contain flag information about every pixel on the sky.

We analyzed a large sample of galaxies from the photometric observations of the Sloan Digital Sky Survey. The data extend over an area of about 3000 square degrees, organized in long, 2.5 degree wide 'stripes'. The stripes are organized into 12 'camcols', corresponding to the detector camera columns, and those are broken up into 'fields' that are pipeline processing units. We downloaded about 50 million galaxies from the project database at Fermilab, and created a custom database of this downloaded data, using Microsoft SQL Server. Each galaxy in the database had a five-color photometry, and an angular position, plus a description of which stripe, camcol, and field it belongs to. Next, we computed the photometric redshifts, absolute luminosities, rest-frame types, and their covariances for each of the 50 million galaxies. These derived data were also inserted into the database. Using this database, we can write SQL queries that generate a list of objects that satisfy a given selection criterion and that are in the angular statistical sample of galaxies. We can also write SQL to get a list of masks, with their rectangular boundaries.

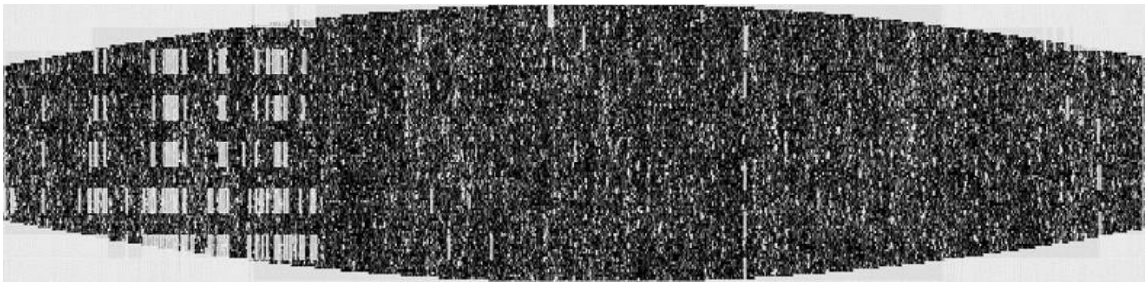

**Figure 1.** The layout of a single 'stripe' of galaxy data in the SDSS angular catalog, with the mask overlay. The stave-like shape of the stripe is due to stripe layout over the sphere. The vertical direction is stretched considerably.

The selection criteria for the redshift survey were much more complex, they involve observing objects selected in 2.5 degree stripes, then observed spectroscopically with a plug-plate of 3 degrees diameter. The centers of the plates were selected to overlap in higher density areas, since optical fibers cannot be placed closer than 55" in a given observation. This complex pattern of intersections and our sampling efficiency in each of the resulting 'sectors' was calculated by hand using spherical polygons (Blanton & Tegmark 2001). We are automating this to be done inside the database, and updated regularly. To cope with this problem, we created a separate database for the redshift-space analysis that has both our statistical sample and a Monte-Carlo simulation of randomly distributed objects that were generated per our angular sampling rate. The size of this latter data set is about 100 million points.

## 4. ANALYSIS OF CLUSTERING

### 4.1 Angular correlations

Starting with 50 million galaxies, we computed the photometric redshifts and inserted the data into the database. Then we started a stepwise sample selection process, first creating volume limited sub-samples, then selecting by luminosity, then by rest frame color. Figure 2 illustrates this selection process. The sub-samples were never physically extracted, only defined in terms of database queries.

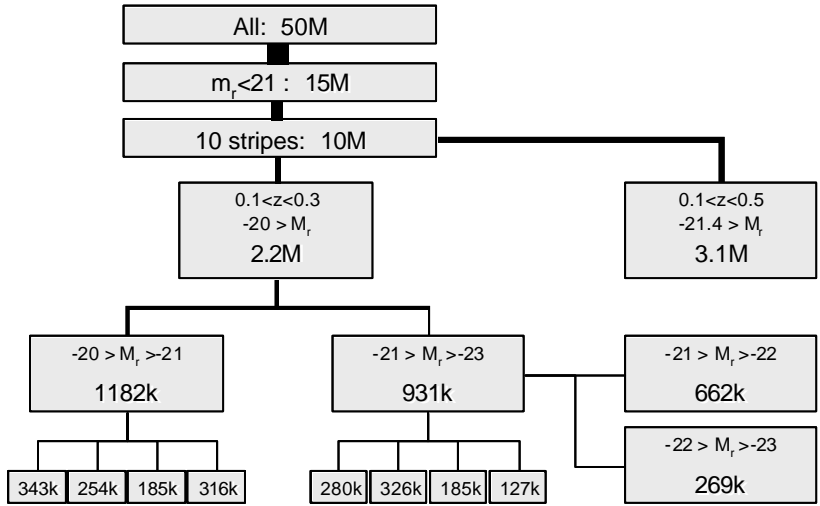

**Figure 2:** The filtering sequence of data for the angular correlation function analysis. Starting with over 50 million galaxies having angular positions and photometric redshifts in the database, we first filtered on quality/brightness, then filtered contiguous regions on the sky. Once this sample was selected, we created volume-limited subsets, and rest-frame color cuts. The samples are virtual data that is only extracted on demand at the time of the analysis.

We used a customized version of eSPICE, a Euclidian version of a fast two-point correlation code developed by I.Szapudi, S. Colombi and S. Prunet to analyze the data stripe by stripe. Within a stripe the Euclidian distance approximation is good enough, since the width of the stripe never exceeds 2.5 degrees. The code based on Fast Fourier Transforms generates a two-dimensional angular correlation function. It first creates a high-resolution two-dimensional rectangular grid, with 1 arcmin cells. It then maps both the galaxy sample and the masks onto the grid. Then it computes the various bin-counts for the windowed variant of the LS estimator using the FFTW package. The correlation package talks directly to the database, it requires a query with the stripe number, and the ranges in redshift, spectral type and luminosity, and it returns the angular correlation function.

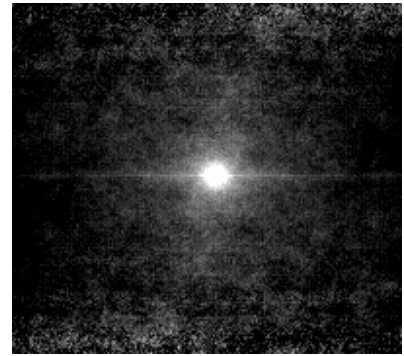

**Figure 3.** The two-dimensional angular correlation function. Note the extended feature along the scan direction at zero lag, due to flat-field vector in a drift scan.

In the two-dimensional angular correlation function there is an elongated streak at zero lag, along the scan direction (see Figure 3). This is due to the flat-field vector for a drift-scan survey: pixels are averaged during drift scan integration, and then divided by a flat-field vector during processing. Any small errors in the flat-field vector will cause a correlation along the columns in the scan direction. This is a well-known property of drift-scan, thus this part of the data has been censored before azimuthal averaging.

We then computed the one-dimensional angular correlation function separately for each sample in each stripe. Next, we averaged the correlation functions over the stripes. The computation of 20 different subsamples, over 10 stripes each, takes about 30 minutes. The resulting error bars on the correlation functions are stunningly small (Figure 4). For comparison the total number of galaxies in the data is in the millions. Using traditional $N^2$ techniques on this problem would have taken several weeks rather than several hours.

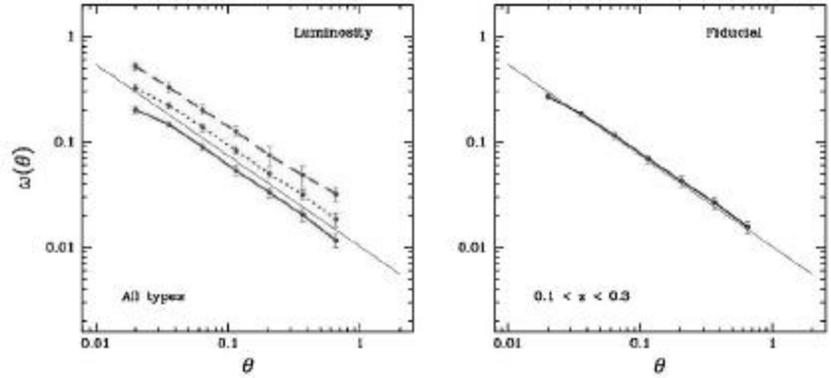

**Figure 4.** The angular correlation function for the SDSS galaxies as a function of luminosity, using photometric redshifts in a volume limited sample. The sample consists of 2.2 million galaxies over 2500 square degrees, the largest sample analyzed to date.

**4.2 Spatial Correlations**

Our spatial correlation analysis used 120 thousand galaxies from the SDSS redshift catalog. This sample currently has rather poor sky coverage. Most of the data is in three main regions, each resembling a 10-15 degree thick slice (Figure 5). Two of the regions are on the Northern and Southern equator, and there is a Northern high latitude region. In the Karhunen-Loeve analysis we currently use these three regions separately. Then, at the end of the analysis we combine their likelihood functions.

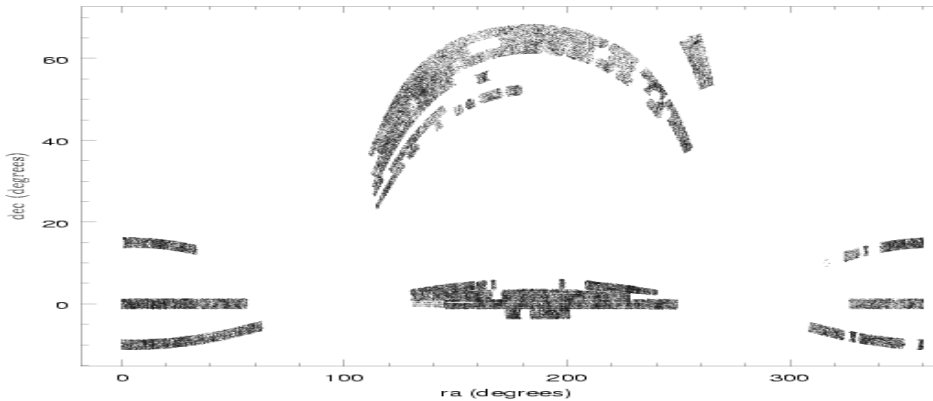

**Figure 5.** The distribution of the SDSS spectroscopic catalog in Sample 9 on the sky. There are 120 thousand redshifts in this catalog. As obvious from the figure, current sky coverage is rather uneven.

Each region is split into approximately 6000 cells. The cells are spherical, and form a dense-packed, hexagonal lattice in three-dimensional space. The geometry of the centers is stored in the database. The database also contains the galaxy catalog and the Monte Carlo simulation of our selections on the sky, using a uniform underlying distribution, and weights indicating fractional completeness on the sky at that point.

Next, we count the galaxies in the cells to estimate of the local overdensity. This is the beginning of the database work. We have three counts to compute: (i) the number of observed galaxies in a cell, (ii) the number of galaxies we would have observed at that distance if we had perfect sky coverage and a perfectly known radial selection function, and (iii) the number of galaxies would we have observed with a perfect radial selection, but with our known incomplete sky coverage. All counting is done inside the database, on the actual redshift sample, and on the preloaded Monte-Carlo sample, with its 100 million objects. The HTM library helps identify all galaxies in a given cone. A simple radial index accelerates counting. The ratios of these counts give two numbers: (1) the relative overdensity of galaxies in the cell, and (2) the effective completeness, or sampling rate of the cell. We reject cells with completeness less than 0.75, since inversion from the incomplete observed galaxy counts to the true counts would be too large. Using this information we can compute a data vector, the galaxy overdensity in the remaining cells, and a shot noise matrix.

Once the geometry is decided, we compute a correlation matrix among the galaxy overdensities for a given fiducial power spectrum, including a shot noise term. Then the 6000x6000 correlation matrix is diagonalized, and all eigenvalues and eigenmodes computed. We sort those by decreasing eigenvalues, and typically discard the 4000 lowest ranked eigenmodes. These mostly contain only shot-noise, while the cosmological signal we are trying to measure is mostly in the first 2000 modes. Figure 6 shows some preliminary results.

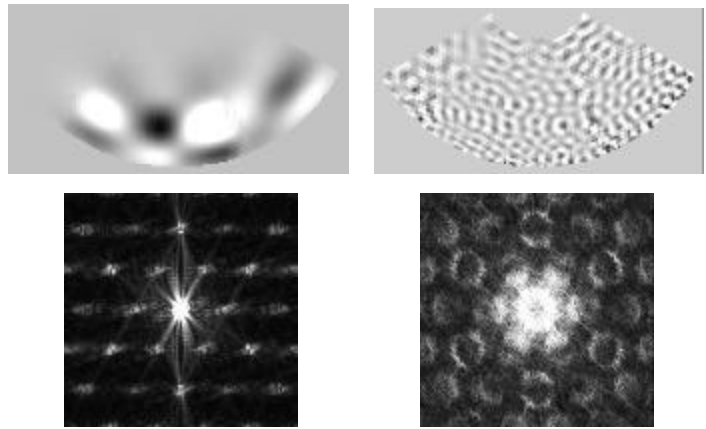

**Figure 6**. Some of the eigenmodes for one of the equatorial regions. The upper half shows the eigenmodes in real space, while the lower half is in Fourier space. The patterns reflect the hexagonal lattice.

At this point we are ready to start computing the likelihood. We first transform the data vector into the KL-transform space, by calculating the dot product with each of the basis vectors. We only keep the first 2000 values. For each new combination of cosmological parameters we compute a new correlation matrix. We then project it down to the selected 2000 dimensional subspace, where the likelihood computations are done. The log likelihood value is given by the scalar formed by sandwiching the inverse of the projected correlation matrix with the data vector and its transpose.

Needless to say, these computations are rather demanding numerically. Since they involve large matrices they require large memory. Hewlett-Packard has given us two computers, one with 28GB, another with 6GB of memory for these calculations. The results about the cosmological parameters are in excellent agreement with results from the Cosmic Microwave Background (deBernardis 2001) and other redshift surveys, like the 2dF (Percival et al 2001). While earlier likelihood calculations took close to a month, now we can do these in little over 30 CPU hours. The emerging bottleneck in the calculations is the fact that we need to address systematic errors, like photometric zero-points, or uncertainties in the radial selection function on the analysis.

## 4.3 Including systematic effects

The most significant uncertainty in our analysis comes from systematic effects of sample selection. The SDSS spectroscopic targets have been selected from the photometric catalog, which is observing objects in stripes, runs, and camera columns. There is a common photometric zero point for each run, and there is a separate flat field vector for each camcol. The current uncertainties in each of these quantities are probably less than 0.02 magnitudes. Taking a random Gaussian with a variance of about 0.015 is probably a good approximation. A small shift in zero point offsets every galaxy magnitude in the respective area. For most of the sample this is a small effect, but at the outer edge of the flux limited survey the selection probability is rapidly decreasing, since we are only selecting the brightest galaxies. Even a small zero point shift at those distances can be amplified dramatically, by a factor of 5-10, due to the steep slope of the selection function, therefore generating an artificial large-scale clustering pattern on the galaxy distribution.

One can approach the issue of systematic errors on two levels. First, we need to be able to simulate this effect, and add such synthetic errors to our data, then redo our analysis and monitor the deviations it causes. This will increase the uncertainties beyond the statistical errors of the hypothesis testing. Secondly, we can desensitize our analysis to these errors, by building an empirical contribution to the correlation matrix arising from the zero-point effects. We are building a database to accomplish both these tasks as follows.

We first generate 100 million random points, uniformly distributed, with a generous overlap, over the current sample area. We assign each of the random points to a stripe, run and camcol. There is a unique mapping from right ascension and declination to these coordinates. Then we calculate a field number and a CCD row and column to each object. Each object also carries a weight, between 0 and 1, which describes the angular selection pattern at that point.

We then generate a table that stores the mock survey layout with stripes, runs and camera columns. For each unique combination we have a placeholder for a zero-point error, which we fill with appropriate random numbers for each realization. By joining the random points table with this zero-point table, we can assign a random weight for each object whether it would have been included in any angular selection (like different subsamples in the angular correlation analysis), by using an appropriate function.

For the three-dimensional analysis, we generate a table with the value of the differential and cumulative selection function as a function of radial distance. This table also contains the logarithmic derivative of the selection function, the modulation coefficient of the zero-point shifts. The modulation coefficient, multiplied with the signed zero-point shift will tell us the change in the selection probability at that distance. Given a value of a zero-point shift, we can use this table to generate a random distance for each galaxy in the sample, reflecting a proper radial distribution and the expected change in selection.

Next, we perform the cell counting from the KL geometry, and create a modulated data vector. This needs to be compared to the unperturbed data vector, and the difference is the modulation signal that needs to be added to our real galaxy counts for a 'mock' realization. It is better to perturb the real counts than just the Monte-Carlo data only, since those have zero clustering signals.

One can do much better by taking the modulated data vector, and create a buffer for the correlation matrix in the database. For a 10,000x10,000 sample this is a table of 50 million entries, containing a row, column index and a value. We add to the correlation matrix element the appropriate product of the modulated vector elements. Repeating this about a hundred times gives a reasonably good ensemble average of the correlation matrix of the zero-point errors. Adding this matrix to the fiducial correlation matrix when building the KL basis should produce a few hundred modes that will reflect the structure of these errors. The number comes from the degrees of freedom: we can assign a separate zero point for each of the 12 camera columns in each stripe. Rejecting those modes from the likelihood analysis will effectively filter out the patterns consistent with zero-point errors from the data, and therefore decrease the sensitivity of the analysis to such errors. This can also be tested and quantified: adding such mock errors before and after the inclusion of the filtering scheme should show a clear difference in the detected power spectrum.

Such an error treatment will be especially be important for analyzing the spatial clustering of the SDSS LRG galaxy sample (Large Red Galaxies) which form an almost volume limited sample out to a redshift of 0.45, covering a volume of over 2 Gpc$^3$. This sample has been selected based upon a complex color cut, selecting the tail of a multidimensional distribution, thus the modulation effects are more complex and much stronger. At the same time it may also be more rewarding. Due to the large volume, the cosmic variance is much smaller, and due its greater depth the curvature of the universe can also be measured, providing a possibility to measure the amount of 'dark energy' in the Universe, one of the greatest challenges in today's cosmology. It would not be possible to attempt this without the complex combination of our analysis tools, large databases and cutting edge computers. This framework took us several years to build, but it is already making an enormous impact as we are doing the analyses.

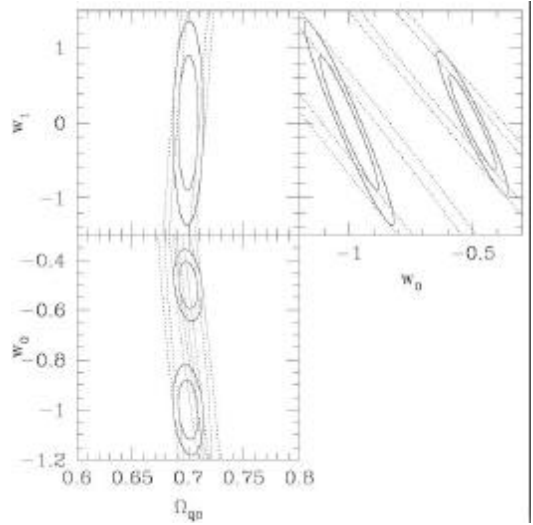

**Figure 7.** The projected errors in determining the equation of state ($w_0$) for the dark energy, and its derivative ($w_1$), using the SDSS LRG sample, using the KL correlation matrix. For a cosmological constant the values would be –1 and 0 (from Matsubara and Szalay 2002).

## 5. SUMMARY

Exponentially growing astronomy data volumes prose serious new problems. Most statistical techniques labeled as `optimal' are based on several assumptions that were correct in the past, but are no longer valid. Namely, the dominant contribution to the variance is no longer statistical – it is systematics, computational resources cannot handle $N^2$ and $N^3$ algorithms -- N has grown from $10^3$ to $10^9$, and cosmic variance can no longer be ignored.

What are the possibilities? We believe one answer lies in clever data structures, borrowed from computer science to pre-organize our data into a tree-hierarchy, and having the computational cost dominated by the cost of sorting, an *NlogN* process. This is the approach taken by A. Moore and collaborators in their tree-code (Moore et al 2001).

Another approach is to use approximate statistics, as advocated by Szapudi et al (2001). In the presence of a cosmic variance, an algorithm that spends an enormous amount of CPU time to minimize the statistical variance to a level substantially below the cosmic variance can be very wasteful. One can define a cost function that includes all terms in the variance and a computational cost, as a function of the accuracy of the estimator. Minimizing this cost-function will give the best possible results, given the nature of the data and our finite computational resources. We expect to see more and more of these algorithms emerging. One nice example of these ideas is the fast CMB analysis developed by Szapudi et al (2002), which reduces the computations for a survey of the size of Planck from 10 million years to approximately 1 day!

In summarizing, several important new trends are apparent in modern cosmology and astrophysics: data volume is doubling every year, the data is well understood, and much of the low level processing is already done by the time the data is published. This makes it much easier to perform additional statistical analyses. At the same time many of the outstanding problems in cosmology are inherently statistical, either studying the distributions of typical objects (in parametric or non-parametric fashion) or finding the atypical objects: extremes and/or outliers. Many of traditional statistical algorithms are infeasible because they scale as polynomials of the size of the data. Today, we find that more and statistical tools use advanced data structures and/or approximate techniques to achieve fast computability.

In both applications we presented, the databases and the computations performed inside were an essential component of the analysis and enabled us to deal with much larger datasets. We also integrated some of our tools with the database itself: like generating plots of galaxy surface densities or the whole angular correlation code itself.

In the not too distant future, when our data sets grow another order of magnitude growth, only *NlogN* algorithms will remain feasible—the cost of computation will become a very important ingredient of an optimal algorithm. Such an evolution in our approach to astrostatistics can only be accomplished with an active and intense collaboration of astronomers, statisticians and computer scientists.

## ACKNOWLEDGEMENTS

TM acknowledges support from the Ministry of Education, Culture, Sports, Science, and Technology, Grant-in-Aid for Encouragement of Young Scientists, 13740150, 2001. AS acknowledges support from grants NSF AST-9802 980, NSF KDI/PHY-9980044, NASA LTSA NAG-53503 and NASA NAG-58590. He is also supported by a grant from Microsoft Research. Our effort has also been generously supported with hardware by Compaq, Intel and HP.